\documentclass[
  ,draft            % use draft while you are working on the paper
  ]
  {aipproc}

\layoutstyle{6x9}

\begin{document}

\title{Effective hamiltonian approach and the lattice  fixed node approximation}

\author{Sandro Sorella and Seiji Yunoki}{
  address={INFM-Democritos National Simulation Centre and SISSA, Via Beirut n.2 ,34014 Trieste, Italy}
}

%%\author{<author2>}{
%%  address={<common address for author2 and author3>}
%%}

%%\author{<author3>}{
%%  address={<common address for author2 and author3>}
%%  ,altaddress={<author1 address>} % additional visiting address
%%}

\begin{abstract}
We define a numerical scheme  that allows to approximate a given Hamiltonian
by an effective one,
by requiring several constraints determined by
exact properties of generic ''short range'' Hamiltonians.
In this way the standard lattice fixed node is also improved as far as 
the variational energy is concerned. 
The effective Hamiltonian is defined in terms of a guiding function
$\psi_G$ and can be solved exactly
by Quantum Monte Carlo methods. We argue  that,  for reasonable $\psi_G$
and away from phase transitions,  the long distance,
low energy properties  are rather independent on the chosen guiding
function, thus allowing to remove the well known problem of  standard
variational Monte Carlo schemes
 based only on total energy minimizations,
and   therefore insensitive to long distance low energy properties.
\end{abstract}

\maketitle

%%%%%%%%%%%%%%%%%%%%%%%%%%%%%%%%%%%%%%%%%%%%
%% MAINMATTER
%%%%%%%%%%%%%%%%%%%%%%%%%%%%%%%%%%%%%%%%%%%%

\section{Introduction}
After many years of intense numerical and theoretical efforts
the problem of strong correlation in 2d or higher dimensional systems is
still open. The main difficulty is to calculate the ground state of a many-body strongly correlated Hamiltonian with a technique which is systematically
convergent to the exact solution with a reasonable computational effort.
 Quite generally
 all the known approximate techniques rely on the variational
principle. The many-electron wavefunction  is determined
 by   an appropriate
 minimization of the energy within a particular class of wavefunctions.
 The Hartree-Fock method is the first  clear  example: here the many-electron wavefunction is approximated by a single Slater determinant. Indeed  also a very
recent technique like the Density-Matrix Renormalization Group
(DMRG)\cite{dmrg} falls in this
class, being certainly a variational approach,
 based on  a particularly smart iteration scheme to define
a  variational wavefunction very good for  low dimensional systems.
However,  within the variational approach,  one faces the following problem:
By increasing the  system size the gap to the first excited state scales
 generally to zero quite rapidly.
 Thus   between the ground state energy and the variational energy there may 
be a very large number
of states with completely different correlation functions.
In this way one can generally obtain different variational wavefunctions
with almost similar
energy, but with completely different correlation functions and therefore 
compelling physical meaning.
By the above consideration it is  easily understood that,
 within a straightforward variational technique and limited accuracy in energy
-say $1\%$, there is no hope to obtain sensible results for large system size,
 unless for model Hamiltonians  with a  finite gap to all
excitations, such as  the simplest  band insulators.
The most striking example of this limitation of the variational approach
is given by the Heisenberg model $H=J \sum_{<i,j>} \vec S_i \cdot \vec S_j$
where it was shown in\cite{doucot} that two wavefunctions  with completely
different long-distance properties, with or without antiferromagnetic long
range order, provide almost similar (and very accurate within $0.1\%$ accuracy)
 energy per site in the thermodynamic limit.

In the following we will consider a possibility to overcome the above
limitation by means of the ''effective Hamiltonian'' approach.
The main task is not to approximate a wavefunction
as in the variational approach,
but more conveniently our effort is to approximate the Hamiltonian $H$
as closely as possible by means of a  correlated Hamiltonian $H^{eff}$
that can be solved numerically by Quantum Monte Carlo schemes.
The important point is that, within this construction,
some important properties of physical short range Hamiltonians are preserved,
providing in this way
  a much better control of correlation functions.

\section{Hamiltonian as matrix elements: back to Heisenberg }

Let us consider 
the configuration basis $\left\{ x \right\}$, where all the $N$ electrons have
definite spin ($\uparrow$ or $\downarrow$)  and positions on a lattice with
$L$ number of sites. 
 The matrix elements
 of an  Hamiltonian   $\bar H$
 in this physical basis will be
indicated by $\bar H_{x,x^\prime}$. Obviously the chosen basis is crucial 
to define the concept of locality, a property of the hamiltonian. 
A physical short range Hamiltonian $\bar H$
has non zero off-diagonal matrix elements $\bar H_{x,x^\prime}$ only  for
configurations $x$ and $x^\prime$  differing one another by
local short-range moves of  electrons, more precisely:
\begin{equation}\label{srcondition}
\bar H_{x^\prime,x} \ne 0 ~~~~~{\rm  if}~~~~ |x-x^\prime| \le  \Lambda
\end{equation}
where $|x-x^\prime|$ indicates  the distance in the $d\times N$ dimensional
space, and $\Lambda << L $ is a suitable constant  denoting the short-range
character of the Hamiltonian $\bar H$.
In this definition the diagonal matrix elements do not play any role,
so not only conventional Hubbard-Heisenberg-t-J model are short range
Hamiltonian (with $\Lambda =1$), but also
models with long range   interactions,
provided these interactions-like the Coulomb one-are
 defined in the basis of configurations $x$, thus representing classical
interactions in absence of  the kinetic term.
 We believe that within this definition, essentially all physical Hamiltonian
can be considered to belong to this class.

\subsection{The  $J_1-J_2$ model}

The simplest     model that describes  frustration of
antiferromagnetism  is the  Heisenberg model
with superexchange couplings extended up to nearest ($J_1$)  and next nearest
neighbor ($J_2$) couplings:
\begin{equation} \label{model}
H = J_{1} \sum\limits_{<i,j>_{n.n.}}  \vec S_i \cdot \vec S_j
 + J_{2} \sum\limits_{<i,j>_{n.n.n}}   \vec S_i \cdot \vec S_j
\end{equation}
where summations $<i,j>_{n.n.}$ (  $<i,j>_{n.n.n.}$  )
 are over the nearest neighbor (next nearest neighbor) lattice sites $R_i,R_j$
and periodic boundary conditions (PBC) are assumed.

Whenever the next-nearest neighbor exchange $J_2$
is large enough compared to the nearest neighbor one $J_1$,
it is widely believed  that  the antiferromagnetic phase
is destabilized, until  a second order transition  takes place 
and  a   phase with a spin  gap and a finite correlation length appears 
for $J_2$ large enough.

\section{The effective Hamiltonian}

We will  define here  a simpler  effective Hamiltonian matrix $H^{eff}$
closely related to $H$, by means of the matrix elements
 $H^{eff}_{x^\prime,x}$
 in the basis $\{x \}$ of configurations
 where all electrons have a definite spin $\uparrow$ or $\downarrow$
 in all lattice sites $R_i$.
 Such an extension of the Hamiltonian $H$, whose matrix elements 
are analogously denoted by $H_{x^\prime,x}$,
  is obtained by means  of the  so called
guiding function $\psi_G(x)$. This wavefunction is  required to be
 non zero for  all configurations $x$.
Once the guiding function is defined for given $\{J_i\}$ the
model $H^{eff}$ can be solved exactly and, as we will show in some
simple case, the low energy properties are  independent of the
low energy properties of $\psi_G$.
The effective Hamiltonian approach  allows to obtain ground state (GS) 
  wavefunctions with non
trivial signs (the one of $\psi_G$),
in this sense representing a more generic GS
of strongly correlated  models.
For instance  the spin  Hamiltonians  that can be solved exactly by
QMC methods
 are the ones for which:
\begin{equation} \label{nosignproblem}
s_{x^\prime,x} = \psi_G(x^\prime) H_{x^\prime,x} \psi_G(x) \le 0
\end{equation}
 for particularly simple $\psi_G(x)$ satisfying the Marshall sign
rule
$$\psi_G(x) \propto  (-1)^{ \rm Number~of~spin~down~in~one~sublattice}.$$
This is the case for the Heisenberg model 1d (gapless), 2chains (gapped
but not spin liquid), 2d (gapless antiferromagnet), where it is also
clear that with the same sign of the wavefunction different
low energy properties  can be obtained by solving exactly $H$ or
 $H^{eff}=H$  being an exact equality in these  simple cases.

Though there are particular models where the Marshall sign and
(\ref{nosignproblem})  are
satisfied even in presence of strong frustration\cite{santoro,sandvick},
it is clear that these  are just particular and not generic models, since
  Eq.(\ref{nosignproblem}) is generally  violated even when
 the GS of $H$ is used in Eq.(\ref{nosignproblem}).
The reason  is  that
for generic frustrated Hamiltonian (with sign problem) there are off
diagonal matrix elements  with $s_{x,x^\prime} >0$, namely some matrix
elements do not decrease the expectation value of the
energy: they are ''unhappy'' even in the ground state as can be simply tested
in the $J_1-J_2$ model for $J_2 \ne  0$ or in even simpler model.

In this case the effective Hamiltonian $H^{eff}$  is defined in terms of
the matrix elements of $H$, 
in order to generate a dynamic as close as possible to the exact one.
An obvious  condition to  require, is that if $\psi_G$ is exact
the ground state of $H^{eff}$ has to coincide with the one of $H$.
In order to fulfill this condition the so called lattice fixed node was
proposed\cite{ceperley}, $H^{eff}=H^{FN}$, and $H^{eff}$ was obtained
by strict analogy with the continuous fixed node scheme.
In the following we will argue that there is a better way to choose the
effective Hamiltonian, which not only provides better variational energies,
but also allows a better accuracy  of low energy long distance properties
of the ground state.
In the standard fixed node approach all the matrix elements that
satisfy Eq.(\ref{nosignproblem}) are unchanged,
 whereas the remaining off-diagonal matrix elements
are dealt semiclassically and traced to the diagonal term of
$H^{eff}_{x,x}$.
The  FN-effective
hamiltonian can be obtained by modifying the diagonal term $H^{eff}_{x,x}$,
in order to have the same local energy of the exact Hamiltonian
for any configuration $x$, namely:
$e_{H}(x)=e_{H^{eff}}(x)$,
where the local energy is defined in terms
of the guiding function $\psi_G$ and an Hamiltonian $\bar H$ by:
\begin{equation} \label{local}
e_{\bar H}(x)=\sum_{x^\prime} \psi_G(x^\prime) \bar H_{x^\prime,x}/\psi_G(x)
\end{equation}
This approach was inspired from the similarity of the
fixed node on continuous systems, and indeed is a well established approach
giving also variational upper bounds of the ground state energy\cite{ceperley}.
However in the lattice case there is an important difference.

Even for the fixed node ground state the number of matrix elements
that do not satisfy the condition (\ref{nosignproblem}) may be a relevant
fraction of the total number of matrix elements, whereas in the continuous
case the so called nodal surface (the analogous of this frustrating
matrix elements) represents just an irrelevant ''surface'' of the phase space.

In order to compensate for this  bias in the dynamic,   here we propose to   
modify slightly the fixed node scheme on a lattice, by compensating this
error in the   diffusion of the electrons:
\begin{equation}
H^{eff}= \left\{ \begin{array}{lrl}
  K  H_{x^\prime,x}  ~~{\rm if}~x^\prime \ne x   &{\rm and }~~  s_{x^\prime,x} <& 0   \\
   ~~0   \,~~~~~{\rm  if}~x^\prime \ne x   & {\rm and }~~  s_{x^\prime,x}  >&0
\end{array}  \right.
\label{gammaham}
\end{equation}
where $K$ is a constant that can be determined in a way that the ground state 
of $H^{eff}$ has the lowest possible expectation value of the energy on 
the exact Hamiltonian $H$. 
This procedure has been attempted previously but is very computer and 
time demanding, so its practical implementation is difficult\cite{effective}.

\subsection{The diffusion constant $K$ and the Lieb-Schultz-Mattis theorem}
In order to determine efficiently 
 the value of the constant $K$ we use a
relation  which is well known in the
continuous fixed node\cite{reynolds} and was used
to correct efficiently the error due to the finite time slice discretization
of the diffusion process.\cite{reynolds}
The method uses that, for small imaginary time ($\Delta \tau$)
, the electron positions  change by means of  the
exact  Hamiltonian  propagation
$\psi_G \to exp( -  H  \Delta t )  \psi_G$,
with a diffusion coefficient determined
 only by  the free  Kinetic operator
(the analogous of the  off-diagonal matrix elements of a lattice Hamiltonian). 
It is possible then to correct  the approximate finite $\Delta t$ dynamic,
by requiring that  it  satisfies  exactly this short time condition, that
mathematically can be simply written as:
\begin{equation} \label{diffusion}
[ \vec  x, [H ,\vec x ] ] = D 
\end{equation}
 where $D=3 \hbar^2/m$
is the diffusion coefficient, $\vec x$ is the electron position operator, 
and $m$ the electron  mass.

In a lattice case, or more generally for a system with  periodic boundary conditions, the lattice position operator $\vec x$ is not well defined,
as it cannot be matched with the boundary conditions, namely  the same
 lattice point with $(x,y)$
 and $(x+L,y)$ coordinates,  related by PBC in a $L \times L$ lattice,
have different values  for $\vec x$.
  Analogously to the Berry's  phase calculation\cite{resta}
the  spin and charge position operators  are
more appropriately defined in the exponential form:
\begin{eqnarray} \label{posmu}
O_{\rho,\mu} (x)&=&  exp( i \sum_R ( \tau_{\mu} \cdot R) \, n_R )  \\
O_{\sigma,\mu} (x) &=& exp( i \sum_R ( \tau_{\mu} \cdot R)  \,S^z_R ) 
\end{eqnarray}
where $\mu=x,y,\cdots$ labels the spatial coordinates, e.g.
$\tau_x= ( 2 \pi/L,0)$, $\tau_y = (0, 2 \pi/L)$ for a $L \times L$
square lattice.  Both operators are defined in the basis of
configurations $x$, as the analogous $\vec x$ does in the continuous case.

  Remarkably the spin position operator $O_{\sigma,\mu}$
is exactly equivalent to the well known Lieb-Schultz-Mattis operator,
used to show a well known properties on the low energy spectrum of spin
one-half Heisenberg Hamiltonians.\cite{lsm}
For a generic spin-$1\over 2$ Hamiltonian
 there may be two independent coupling constants
 $K_{\rho}, K_{\sigma}$
 that can  be used to rescale the off-diagonal matrix elements
 and correct the spin and charge lattice diffusion constants independently.
For instance in the $t-J$ model
the charge diffusion is determined  by the hopping matrix elements proportional
to $t$ and the spin-diffusion is set by the $J$ matrix elements.

After simple inspection the following relation holds
both for $O_{\mu,\sigma}$ and $O_{\rho,\sigma}$ (thus we omit
$\sigma,\rho$ labels):
\begin{eqnarray} \label{relation}
& & \langle \psi_G | \left[ O^{\dag}_{\mu}  ,\left[  \bar H , O_{\mu} \right]
 \right]
 |\psi_G \rangle  \nonumber  =\\
& &-\sum_{x\ne x^\prime}  \psi_G(x) \psi_G(x^\prime)
\bar H_{x,x^\prime} |O_{\mu}(x)-O_{\mu} (x^\prime) |^2
\end{eqnarray}
 This quantity can be very simply calculated by standard variational
Monte Carlo both for $\bar H=H$ and $\bar H= H^{eff}$, at fixed guiding
function $\psi_G$.
In this way the value of the undetermined constant $K_{\sigma}$ ($K_{\rho}$)
is very well determined with high degree of statistical accuracy by imposing
that both the effective Hamiltonian and the exact one have the same
expectation value for the above quantity.

\subsection{The final scheme}

The constant $K$  and therefore
 the  effective model $H^{eff}$ (\ref{gammaham})  are  
 {\em uniquely} defined in terms of $\psi_G$ and the 
 exact Hamiltonian $H$.  
 The ground state $\psi_0^{eff}$
 and low
energy excitations of $H^{eff}$ can be computed without sign problem.
For a spin Hamiltonian, only the value of $K_\sigma <1 $ is required, whereas
for the Hubbard model only $K_\rho <1 $ is  important.

In order to compute the expectation value of the energy
 $\langle \psi_0^{eff}| H | \psi_0^{eff}  \rangle $ over this
approximate ground state  for $H$    (or at least an
upper bound as in the standard lattice FN), one can use the method described
in\cite{effective}, which typically sizably improves the standard FN
 upper bound even in the standard case with $K_{\sigma}=K_{\rho}=1$.
As remarked in\cite{effective}, it is not true
(as in the continuous case) that in the lattice the lowest variational
energy value correspond to $K=1$.

The clear  advantage of the effective Hamiltonian $H^{eff}$
 is that it remains in
the same physical Hilbert space of $H$ (compare for instance with
large $N$ or infinite dimension schemes) and, if  
  universal low energy properties  for generic model Hamiltonians
are concerned, {\em it is just irrelevant}  that $H^{eff}$ is
slightly  different from $H$.
In fact when we write down a model Hamiltonian in order to understand
low energy properties (such as order, spin gap etc.)  the underlying
assumption is that between similar Hamiltonians (with similar matrix
elements) the low energy properties cannot be too much different.
If this is not the case, it is not even justified to write down
$H$ itself, rather 
 the complete solution of the {\em all} electron problem
with electron-electron and  electron-ion Coulomb interaction  should
be  fully considered: a clearly prohibitive  task so far.

In a lattice case the effective Hamiltonian $H^{eff}$ does not even
imply a restriction of the Hilbert space (as in the continuous case
where fixing the nodes determines a boundary condition that may not be
satisfied by the excitations) and  therefore it represents also a meaningful
approximation to study the properties of its excitation spectrum.
%Considering the above argumentation it is not clear why the fixed
%node effective hamiltonian  has not been  studied extensively in order
%to solve the many unsolved low energy properties of strongly
%correlated electrons.

%%Some url test \url{http://www.world.universe}.

\section{Results and Conclusions}

We have shown in a previous work\cite{caprio}  that, with the simple 
variational method,  it is possible
to obtain an almost exact representation of the GS
wavefunction on small sizes,   up to $6x6$  sites,  
where exact diagonalization is possible.
This remains true  even in the
strongly  frustrated regime where also the Marshall
sign rule is violated. That the correct  signs of the wavefunction
can be obtained with a BCS wavefunction  (an uncorrelated one)
is one of the most important facts  that comes out from exact diagonalization
on small sizes.

We present here preliminary results on the $J_1-J_2$ model, by comparing
the variational approach (VMC) 
the standard Fixed node  method  (FN) and  the proposed one (FNSR) with 
$K_{\sigma} \ne 1$.
As it is seen from Tab.(\ref{table}) 
 the value of $K_{\sigma}$ is sizably different
from zero in the spin liquid region and allows a remarkable improvement
in the variational energy, significantly  closer 
to  the exact results available 
on this small clusters.
 The value of $K_{\sigma}$ can reach values as small  as $0.5$, much different 
 from the standard approach. 

The reason of such a difference from the continuous case is easily understood.
In the lattice case the number of matrix elements that can provide a 
sign change to a given configuration $x$, may be a considerable fraction 
of all the possible ones. For instance if we take for $\psi_G$ a guiding 
function with the Marshall sign and consider $J_2>0$, all the spin-flip 
matrix elements determined by $J_2$ -namely almost half of all possible 
spin-flips-, are removed by the fixed node scheme (\ref{gammaham}). 
In the continuous case instead only the configurations that are on the so 
called  nodal surface (where $\psi_G(x)=0$) 
may be considered in an analogous situation, implying that the short time 
 diffusion (\ref{diffusion}) 
is   exactly satisfied for almost all configurations  $x$, implying $K=1$ 
in the limit when the fixed node is implemented exactly, namely with 
vanishing small  $\Delta t$ time step error.

%\begin{figure}
%  \includegraphics[height=.3\textheight]{golfer}
%  \caption{Picture to fixed height}
%\end{figure}

\begin{table}
\caption {\label{table}  Comparison of energies between standard fixed node
(FN) and the present improved one (FNSR) as a function of $J_2/J_1$}
\begin{tabular}{|c|c|c|c|c|c|}
\hline
   $J_2/J_1$  & VMC   &  FN  &  $K_{\sigma}^{-1}$  &  FNSR  &   Exact  \\
\hline
 0.00   & -0.65112(5)  & -0.6752(5)  & 1  &  -0.6752(5) &  -0.6789   \\
 0.10   & -0.61869(6) & -0.6326(1)  & 1.1093(1) &  -0.6342(2) & -0.6381   \\
 0.20  & -0.58700(4)  & -0.5942(1)  & 1.2365(3) &-0.5962(1)  &  -0.5990 \\
 0.30  & -0.55646(4)  & -0.55937(3) & 1.3831(5) & -0.56063(4)  & -0.5625  \\
 0.40  & -0.52732(1) & -0.52832(2)  & 1.5406(8)  & -0.52891(2) & -0.5297  \\
 0.45 & -0.51372(2) & -0.51441(2) &  1.6146(9)  & -0.51490(2)  &  -0.5157  \\
 0.50  & -0.50117(2) & -0.50203(2)  & 1.677(1)  & -0.50265(2)  &  -0.5038 \\
 0.55  & -0.49024(2)  & -0.49144(2)   & 1.732(1)  & -0.49241(2)  &  -0.4952  \\
\hline
\end{tabular}
\end{table}

As  far as correlation functions are concerned we present
 in Tab. (\ref{table2})
the estimate of the static spin structure factor obtained with standard 
forward walking technique\cite{calandra}:
\begin{equation} \label{spi}
S(\pi,\pi)=\sum_R  e^{ i (\pi,\pi) R } 
\langle \psi_0^{eff} | S^z_0 S^z_R | \psi_0^{eff} \rangle 
\end{equation}
and  using as guiding function 
 the variational wavefunction obtained  in \cite{caprio}.

From the table we see that the value at $J_2=0$  slightly departs 
from the exact value both for the $FN$ and the $FNSR$ technique, which 
in this case should be the same and exact. 
The problem is that the guiding function $\psi_G$ vanishes 
on a small size for  a considerable fraction of  configurations, preventing 
us to obtain the exact result. 
The large number of zero's for $\psi_G$ 
affects   also the small $J_2$ region, where 
 indeed  the  FNSR does not improve the FN technique.
However the situation drastically changes  in the strongly frustrated regime, 
where the number of zero's  is vanishingly small, the wavefunction being 
much more accurate, and the FNSR provides essentially exact results, 
by considerably improving both the standard VMC and FN approaches.

It is clear however that further and more systematic work is necessary to 
clarify the relevance of the proposed method compared with   the 
conventional ones.  Certainly it greatly simplifies -being equivalent in spirit-
 the standard SR technique\cite{effective}, as the latter one may also provide 
even better variational energies, but very similar correlation functions, 
which should represent our main  task in the study of strongly correlated
systems.  
\begin{table}
\caption {\label{table2}  Comparison of the static magnetic
structure factor $S(\pi,\pi)$   between standard fixed node
(FN) and the present improved one (FNSR) as a function of $J_2/J_1$. The values  of $K_{\sigma}$ are the ones of the previous table.}
\begin{tabular}{|c|c|c|c|c|}
\hline
   $J_2/J_1$  & VMC   &  FN  &   FNSR  &   Exact  \\
\hline
 0.00  & 1.903(4)  & 3.06(13)  & 3.06(13)  &  2.518     \\
 0.10  & 1.840(8)  & 3.27(2)   & 2.94(9)   &    \\
 0.20  & 1.733(7)  & 2.86(1)   & 2.94(2)   & 2.2295   \\
 0.30  & 1.645(7)  & 2.26(1)   & 2.47(1)   & 2.0132 \\
 0.40  & 1.505(7)  & 1.687(7)  & 1.766(7)  &  1.6604  \\
 0.45  & 1.394(6)  & 1.430(5)  & 1.439(7)  & 1.4309   \\
 0.50  & 1.258(5)  & 1.214(5)  & 1.167(5)  & 1.1695   \\
 0.55  & 1.124(5)  & 1.012(4)  & 0.927(5)  &  0.8946  \\
\hline
\end{tabular}
\end{table}

\begin{theacknowledgments}
This work was partially supported by Miur Cofin-2001 and by INFM-PAIS-MALODI.
We thank F. Becca for providing us unpublished results on the 
 $J_1-J_2$ model on a $6\times6$ lattice. 
\end{theacknowledgments}

%% If the bibliography is produced without BibTeX comment out the


\end{document}